\documentclass[conference,letterpaper]{IEEEtran}

\usepackage[utf8]{inputenc}
\usepackage[T1]{fontenc}
\usepackage{url}
\usepackage{ifthen}
\usepackage[compress]{cite}
\usepackage[cmex10]{amsmath}

\usepackage{algorithmic}
\usepackage{graphicx}
\usepackage{textcomp}
\usepackage{xcolor}

\usepackage{amssymb}
\usepackage{amsfonts}
\usepackage{mathtools}
\usepackage{enumerate}

\usepackage{physics}
\usepackage{relsize}

\usepackage{tikz}
\usepackage{pgfplots}
\pgfplotsset{scaled y ticks=false}
\usetikzlibrary{arrows,shapes,backgrounds,plotmarks,positioning}
\pgfplotsset{compat=newest}                    
\pgfplotsset{plot coordinates/math parser=false}
\newlength\figureheight
\newlength\figurewidth

\allowdisplaybreaks[4]

\usepackage{geometry}
\newtheorem{theorem}{Theorem}

\newtheorem{proposition}{Proposition}

\newtheorem{remark}{Remark}

\DeclareMathOperator*{\argmax}{arg\,max}

\newcommand{\Set}[1]{\{#1\}}

\newcommand{\X}{\mathcal{X}}
\newcommand{\Y}{\mathcal{Y}}

\newcommand{\Leak}{\mathcal{L}}

\newcommand{\XEntropy}{H}

\newcommand{\E}{\mathbb{E}}

\newcommand{\supp}{\text{supp}}

\allowdisplaybreaks

\begin{document}
\title{$\alpha$-leakage Interpretation of R\'{e}nyi Capacity}
\author{%
	\IEEEauthorblockN{Ni Ding}
	\IEEEauthorblockA{University of Auckland \\
		New Zealand \\
		ni.ding@auckland.ac.nz}
	\and
	\IEEEauthorblockN{Farhad Farokhi}
	\IEEEauthorblockA{University of Melbourne \\
		Australia \\
		farhad.farokhi@unimelb.edu.au}
	\and
	\IEEEauthorblockN{Tao Guo, Yinfei Xu and Xiang Zhang}
	\IEEEauthorblockA{Southeast University\\
		China\\
		\{taoguo, yinfeixu, xiangzhang369\}@seu.edu.cn}
}

\maketitle

\begin{abstract}
For $\tilde{f}(t) = \exp(\frac{\alpha-1}{\alpha}t)$, this paper shows that the Sibson mutual information is an $\alpha$-leakage averaged over the adversary's $\tilde{f}$-mean relative information gain (on the secret) at elementary event of channel output $Y$ as well as the joint occurrence of elementary channel input $X$ and output $Y$. This interpretation is used to derive a sufficient condition that achieves a $\delta$-approximation of $\epsilon$-upper bounded $\alpha$-leakage. A $Y$-elementary $\alpha$-leakage is proposed, extending the existing pointwise maximal leakage to the overall R\'{e}nyi order range $\alpha \in [0,\infty)$. Maximizing this $Y$-elementary leakage over all attributes $U$ of channel input $X$ gives the R\'{e}nyi divergence. Further, the R\'{e}nyi capacity is interpreted as the maximal $\tilde{f}$-mean information leakage over both the adversary's malicious inference decision and the channel input $X$ (represents the adversary's prior belief). This suggests an alternating max-max implementation of the existing generalized Blahut-Arimoto method.
\end{abstract}

\section{Introduction}

The quantification of privacy leakage was originally proposed in a Bayesian inference setting~\cite{Calmon2012_Allerton} as follows.
Compute the information on the secret corresponding respectively to the prior belief (before the adversary observes channel output $Y$) and the posterior belief (after the adversary observes channel output $Y$). Take the privacy leakage as the increase in the adversary's information on the secret from prior to posterior belief.
Such measure ranges from the mutual information~\cite{Calmon2012_Allerton,Sankar2013_TIFS}, maximal leakage in~\cite{Issa2020_MaxL_JOURNAL} and $\alpha$-leakage~\cite{Liao2019_AlphaLeak}, a R\'{e}nyi measure for order $\alpha\in[1,\infty)$ tunable between mutual information (average leakage) and maximal leakage.
In~\cite{Ding2024_ISIT}, a R\'{e}nyi cross entropy is proposed, extending the definition of $\alpha$-leakage to the entire R\'{e}nyi order range $\alpha\in[0,\infty)$ which well interprets Arimoto mutual information as a privacy risk assessment.

Independently, recent work \cite{Ding2024ITW} proposes another $\alpha$-leakage following the intuitions of R\'{e}nyi measures in \cite{Renyi1961_Measures}.
Instead of separate calculations for prior and posterior, the elementary (instance-wise) relative information gain jointly incurred is obtained, and the $\tilde{f}$-mean\footnote{For invertible and continuous $f$, the $f$-mean of $X$ is $\bar{X} = f^{-1}(\E[f(X)])$, which is also called quasi-arithmetic mean.}
of all elementary measures quantifies the information leakage. Here, $\tilde{f}(t) = \exp(\frac{\alpha-1}{\alpha}t)$.
It is shown in \cite[Theorem~1]{Ding2024ITW} that the maximum of this $\tilde{f}$-mean measure over all adversary's estimation decisions (on the secret) equals to R\'{e}nyi divergence. This, via the expression \cite[Eq.~(20)]{Ding2024_ISIT}, further interprets Sibson mutual information as a $\tilde{f}$-mean of $Y$-elementary leakage, the relative information gain incurred at each instance $y$ of channel output $Y$.

Following \cite{Ding2024ITW}, we reveal in this paper that the Sibson mutual information is also a $\tilde{f}$-mean of $XY$-elementary leakage at the joint
occurrence $(x,y)$ of channel input $X$ and output $Y$.
Using exponential Chebyshev tail bound, we derive a sufficient condition in terms of Sibson mutual information that guarantees a $\delta$-approximation of both $Y$- and $XY$-elementary leakage is upper bounded by $\epsilon$.
We reveal a $\tilde{f}$-mean interpretation of $\alpha$-leakage measure proposed in~\cite[Definition]{Ding2024_ISIT}, \cite{Issa2020_MaxL_JOURNAL,Liao2019_AlphaLeak}, and derive the corresponding $Y$-elementary leakage and show that the maximum of it over all attributes $U$ of channel input $X$ is the $\alpha$-leakage measured by R\'{e}nyi divergence. 
This result extends the existing pointwise maximal leakage (PML)~\cite{Saeidian2023_PML}  to the overall R\'{e}nyi order range $\alpha \in [0,\infty)$,\footnote{In this paper, for R\'{e}nyi order range such as $[0,\infty)$, we keep the $\infty$ side open. But, it should be clear that all results apply to the order $\alpha = \infty$.}
where PML corresponds to the case $\alpha = \infty$.
We further reveal that the R\'{e}nyi capacity $C_\alpha$ is the maximal $\tilde{f}$-mean information leakage over both the adversary's estimation decision on secret and the channel input $X$, which represents the adversary's prior belief.
This parallels the existing definition of R\'{e}nyi capacity as the max-min of conditional R\'{e}nyi divergence and suggests an alternating max-max implementation of the existing generalized Blahut-Arimoto method~\cite{Arimoto1976_CompErrExp} for computing $C_\alpha$.

\subsection{Notation}
Random variables (r.v.) are denoted by capital Roman letters while their elementary events, instances, or realizations are denoted by lowercase Roman letters. Sets are denoted by calligraphic Roman letters. For instance, $x$ in an instance or realization of r.v. $X$ and $\X$ refers to its alphabet. We assume finite countable alphabet in this paper.
Let $P_{X}(x)$ be the probability of outcome $X = x$ and $P_X  = (P_{X}(x) \colon x \in \X)$ be the probability mass function.
The support of $P_{X}$ is defined as $\supp(P_{X}) = \Set{x \colon  P_X(x) > 0}$.
The expected value of $f(X)$ with respect to (w.r.t.) probability mass function $P_{X}$ is  $\E_{P_X}[f(X)] = \sum_{x \in \X} P_{X}(x) f(x)$.
The probability distribution of $Y$ conditioned on the event $X = x$ is denoted by $P_{Y|X=x} = (P_{Y|X}(y|x) \colon y\in\Y)$. 
Furthermore, $P_{Y|X} = (P_{Y|X}(y|x) \colon x\in\X, y\in\Y)$.
For r.v. $X$, we denote $X_{\alpha}$ the $\alpha$-scaled $X$ with probability $P_{X_{\alpha}}(x) = {P_{X}^{\alpha}(x)}/{\sum_{x}P_{X}^{\alpha}(x)}$ for all $x \in \X$.

\section{$\alpha$-Leakage: Maximum Information Gain}
\label{sec:Leak}

Estimation theory can be used to define information leakage~\cite{Calmon2012_Allerton, Issa2020_MaxL_JOURNAL,Liao2019_AlphaLeak,Saeidian2023_PML}. A privacy-preserving channel can be modeled via conditional probability $P_{Y|X}$. Here, the channel output $Y$ can be accessed by all users, including malicious or adversarial entities, while $X$ is the raw input, which may contain sensitive information. An adversary can construct an estimate of $X$, denoted by $\hat{X}$, based on the output $Y$ using conditional probability $P_{\hat{X}|Y}$. 
This induces a Markov chain $X - Y - \hat{X}$.
In this context, the adversary's prior belief about the secrete r.v. $X$ is captured by $P_X$ while the adversary's posterior belief is given by $P_{\hat{X}|Y}$ and can be used to measure information gain.
The adversary will seek the optimal estimation $P_{\hat{X}|Y}^*$ that maximizes the information gain, which incurs the maximal privacy leakage on the channel input $X$.

\subsection{Leakage Measure by Divergence}

For $f(t) = \exp((\alpha-1)t)$, R\'{e}nyi divergence is defined in~\cite{Renyi1961_Measures} as an $f$-mean relative information gain as
\begin{multline}
	D_{\alpha} (P_{X|Y=y} \| P_X ) \\
	= \frac{1}{\alpha-1} \log  \sum_{x} P_{X|Y}(x|y) \exp((\alpha-1) \iota_{P_{X|Y=y} \| P_{X}}(x) ) \nonumber
\end{multline}
where $\iota_{P_{X|Y=y} \| P_{X}}(x) = \log \frac{P_{X|Y}(x|y)}{P_{X}(x)}$ measures the information increase in $P_{X|Y}$ from $P_{X}$ for each elementary event $x$.

Let $P_{X}$ be the reference probability when we measure the information gain. 
A $\tilde{f}$-mean relative information gain measure for $\tilde{f} (t) = \exp(\frac{\alpha-1}{\alpha}t)$ is proposed in~\cite{Ding2024ITW} as follows.
Let
\begin{equation}
	 \iota_{P_{\hat{X}|Y=y} \| P_{X}}(x) = \log \frac{P_{\hat{X}|Y}(x,y)}{P_X(x)}
\end{equation}
be the elementary information gain incurred by the adversary's estimation decision $P_{\hat{X}|Y}$ relative to $P_{X}$. For each $x$, the appearance frequency of $\iota_{P_{\hat{X}|Y=y} \| P_{X}}(x) $ is determined by the posterior probability $P_{X|Y}(x|y)$.
The $\tilde{f}$-mean of  $\iota_{P_{\hat{X}|Y=y} \| P_{X}}$ is 
\begin{align}
	 \tilde{D}_{\alpha} & \big( P_{\hat{X}|Y=y} \|  P_X  | P_{X|Y=y}  \big) \nonumber  \\
	 &= \frac{\alpha}{\alpha-1} \log \sum_{x} P_{X|Y}(x|y)  \exp( \frac{\alpha-1}{\alpha} \iota_{P_{\hat{X}|Y=y} \| P_{X}}(x)) \nonumber \\
	 &= \frac{\alpha}{\alpha-1} \log \sum_{x} P_{X|Y}(x|y)   \Big( \frac{P_{\hat{X}|Y}(x|y)}{P_X(x)}  \Big)^{\frac{\alpha-1}{\alpha}}
\end{align}
for all $\alpha \in (0,1) \cup (1,\infty)$. At extended orders, 
\begin{align*}
 	\tilde{D}_{\alpha} \big( P_{\hat{X}|Y=y}  \|  P_X & | P_{X|Y=y}  \big)= \\
	&
	\begin{cases}
        \log \min\limits_{x} \frac{P_{\hat{X}|Y}(x|y)}{P_X(x)} & \alpha = 0, \\
        \sum\limits_{x} P_{X|Y}(x|y) \log \frac{P_{\hat{X}|Y}(x|y)}{P_X(x)} & \alpha = 1, \\
        \log \sum\limits_{x} P_{X|Y}(x|y) \frac{P_{\hat{X}|Y}(x|y)}{P_X(x)} & \alpha = \infty.
     \end{cases}
\end{align*}
We call $	\tilde{D}_{\alpha} \big( P_{\hat{X}|Y=y} \|  P_X  | P_{X|Y=y}  \big)$ the $\tilde{f}$-mean conditional divergence between $P_{\hat{X}|Y=y} $ and $P_X $ given $P_{X|Y=y}$.
It is proved in \cite[Theorem~1]{Ding2024ITW} that the maximization of this $\tilde{f}$-mean information gain gives the $f$-mean relative information: for all $\alpha \in [0,\infty)$,
\begin{align}
	 & D_\alpha(P_{X|Y=y}  \|  P_X) \nonumber \\
	 &\quad\quad = \max_{ P_{\hat{X}|Y=y} }  \tilde{D}_{\alpha} \big( P_{\hat{X}|Y=y} \|  P_X  | P_{X|Y=y}  \big), \quad  \forall y \label{eq:RenyiLeakge}
\end{align}
with the maximizer
\begin{equation} \label{eq:OptDes}
P_{\hat{X}|Y}^*(x|y) = \frac{P_{X|Y}^{\alpha}(x|y)/P_{X}^{\alpha-1}(x) }{\sum_{x} P_{X|Y}^{\alpha}(x|y)/P_{X}^{\alpha-1}(x) }, \ \forall  x,y.
\end{equation}
There is a proof of \eqref{eq:RenyiLeakge} in~\cite{Ding2024ITW} by convex optimization. We will show another proof by a decomposition of the unconditional $\tilde{f}$-mean relative information measure in Section~\ref{sec:Decompose1}.
The interpretation of \eqref{eq:RenyiLeakge} is that the R\'{e}nyi divergence $D_\alpha(P_{X|Y=y}  \|  P_X)$ quantifies the adversary's maximal information gain on $X$ after observing the channel output outcome $Y=y$ and therefore indicates the maximal leakage from $X$ at each instance $y$.

\subsection{Maximal Leakage by Sibson Mutual Information}
\label{sec:Sibson}
Given channel $P_{Y|X}$ and input distribution $P_{X}$, the Sibson mutual information~\cite{Sibson1969_InfRadius}
\begin{equation*}
	I_{\alpha}^{\text{S}} (P_X,P_{Y|X}) = \frac{\alpha}{\alpha-1} \log \sum_{y \in \Y} \Big(  \sum_{x\in\X} P_{X}(x) P_{Y|X}^\alpha(y|x) \Big)^{\frac{1}{\alpha}}.
\end{equation*}
is originally defined as the information radius of $f$-mean R\'{e}nyi divergence.\footnote{Information radius, as defined in \cite[Section~2]{Sibson1969_InfRadius}, is a probability distribution that minimizes $f$-mean R\'{e}nyi divergence from a given set of probabilities. 
}
We show below that the Sibson mutual information can be viewed as a maximal leakage measure in terms of $\tilde{f}$-mean relative information gain:\footnote{The equation array \eqref{eq:SibsonRenyiDiv} to \eqref{eq:MaxErrExp} was first presented in \cite{Ding2024ITW} without~\eqref{eq:ITWmissing}, which is a necessary step to prove~\eqref{eq:GainSibsonFMean} by separable optimizations. }
for all $\alpha \in [0,\infty)$,
\begin{align}
	&I_{\alpha}^{\text{S}} (P_{X};P_{Y|X}) \nonumber\\
	&= \frac{\alpha}{\alpha-1} \log \E_{P_Y} \big[ \exp \big( \frac{\alpha-1}{\alpha} D_\alpha(P_{X|Y=y}  \|  P_X) \big) \big] \label{eq:SibsonRenyiDiv} \\
	&= \frac{\alpha}{\alpha-1} \log \E_{P_Y} \big[ \exp \big( \frac{\alpha-1}{\alpha} \nonumber \\
	& \qquad\qquad\quad  \max\limits_{ P_{\hat{X}|Y=y}}  \tilde{D}_{\alpha} \big( P_{\hat{X}|Y=y} \|  P_X  | P_{X|Y=y}  \big)\big) \big] \nonumber\\
	& = \begin{cases}
		 - \min\limits_{P_{\hat{X}|Y}} \log  \max\limits_{y} \exp \big(- \\
		 \qquad \tilde{D}_{0} \big( P_{\hat{X}|Y=y} \|  P_X  | P_{X|Y=y}  \big) \big) 	& \alpha = 0\\
        \frac{\alpha}{\alpha-1} \min\limits_{ P_{\hat{X}|Y}} \log  \E_{P_Y} \big[ \exp \big( \frac{\alpha-1}{\alpha}  \\
		 \qquad \tilde{D}_{\alpha} \big( P_{\hat{X}|Y=y} \|  P_X  | P_{X|Y=y}  \big) \big) \big]	& \alpha \in (0,1) \\
		 \max\limits_{ P_{\hat{X}|Y}} \E_{P_Y} \Big[ \tilde{D}_{1} \big( P_{\hat{X}|Y=y} \|  P_X  | P_{X|Y=y}  \big)  \Big]   & \alpha =1 \\
		 \frac{\alpha}{\alpha-1} \max\limits_{ P_{\hat{X}|Y}}  \log  \E_{P_Y} \big[ \exp \big( \frac{\alpha-1}{\alpha}  \\
		 \qquad \tilde{D}_{\alpha} \big( P_{\hat{X}|Y=y} \|  P_X  | P_{X|Y=y}  \big) \big) \big]   & \alpha \in (1,\infty) \\
         \max\limits_{ P_{\hat{X}|Y}} \log   \E_{P_Y} \big[ \exp \big(  \\
		 \qquad \tilde{D}_{\infty} \big( P_{\hat{X}|Y=y} \|  P_X  | P_{X|Y=y}  \big) \big) \big]	& \alpha = \infty\\
	\end{cases} \label{eq:ITWmissing}\\
	&=\max\limits_{ P_{\hat{X}|Y}}   \frac{\alpha}{\alpha-1} \log \E_{P_Y} \big[ \exp \big( \frac{\alpha-1}{\alpha} \nonumber \\
	& \qquad\qquad\quad  \scalebox{1.1}{$\tilde{D}_{\alpha} \big( P_{\hat{X}|Y=y} \|  P_X  | P_{X|Y=y}  \big)\big) \big] $ }  \label{eq:GainSibsonFMean} \\
	& = \max\limits_{ P_{\hat{X}|Y}}  \frac{\alpha}{\alpha-1}  \log    \sum\limits_{x,y} P_{Y|X}(y|x) P_X(x) \Big( \frac{P_{\hat{X}|Y}(x|y)}{P_X(x)}  \Big)^{\frac{\alpha-1}{\alpha}}
	\label{eq:MaxErrExp}
\end{align}
Eq.~\eqref{eq:SibsonRenyiDiv} was first presented in~\cite[Eqs.~(19) and (20)]{Ding2024_ISIT} showing that the Sibson mutual information is the $\tilde{f}$-mean
of R\'{e}nyi divergence.
There is a conversion from minimization to maximization in~\eqref{eq:ITWmissing}. It is because the measure $\exp( \frac{\alpha-1}{\alpha} D_\alpha(P_{X|Y=y}  \|  P_X) )$ changes from information loss to information gain when $\alpha$ increases above $1$. A similar situation can be found in the proof of \cite[Theorem~1]{Ding2024_ISIT}.
Eq.~\eqref{eq:MaxErrExp} actually formulates a $\tilde{f}$-mean of $XY$-elementary leakage, which will be utilized to derive a tail bound for attaining the $\delta$-approximation of $\epsilon$-leakage in Section~\ref{sec:Tail}.

\subsection{Elementary (Pointwise) Information Leakage}
Independently, a $Y$-elementary leakage measure, called \emph{pointwise maximal leakage (PML)},\footnote{The word `pointwise' is used in~\cite{Saeidian2023_PML}, while we use 'elementary' as it refers to the elementary/atomic event in the sample space.} is proposed in~\cite{Saeidian2023_PML}, quantifying the adversary's maximal information gain incurred at each elementary event of channel output $Y$. The idea is based on the definition of information gain in~\cite{Issa2020_MaxL_JOURNAL,Liao2019_AlphaLeak,Saeidian2023_PML} related to Arimoto conditional entropy and Arimoto mutual information~\cite{Arimoto1977}.

While the PML only defines the $Y$-elementary leakage at $\alpha=\infty$. We propose in this section a generalized measure in the overall R\'{e}nyi order range $\alpha \in [0,\infty)$.
This is done by borrowing the definition of R\'{e}nyi cross entropy in~\cite{Ding2024_ISIT}, a measure that interprets Arimoto conditional entropy in privacy leakage, revising the $\alpha$-leakage in~\cite{Liao2019_AlphaLeak} such that it is well defined by the intuition of R\'{e}nyi measures including order $\alpha = 0$.
The resulting measure is a new $Y$-elementary leakage, relates to Arimoto mutual information, but proved to be upper bounded by the R\'{e}nyi divergence $D_{\alpha} (P_{X|Y=y} \| P_X )$. 

For $U$ being the sensitive attribute of channel input $X$, assume the adversary makes a soft decision $P_{\hat{U}|Y}$ to get the estimate $\hat{U}$. The information leakage is quantified on the Markov chain  $U-X-Y-\hat{U}$. It can be quantified by the Arimoto mutual information $I_\alpha^{\text{A}}(P_U,P_{Y|U})$: for all $\alpha \in [0,\infty)$,

\begin{align}	
	\Leak_\alpha (U & \rightarrow Y)  \nonumber \\
	& =  \frac{\alpha}{1-\alpha} \log \E_{P_Y} \Big[  \exp \Big( \frac{1-\alpha}{\alpha} \Big( \min_{P_{\hat{U}}}  \XEntropy_\alpha (P_U, P_{\hat{U}} )  \nonumber \\
	& \qquad  -  \min_{P_{\hat{U}|Y=y}} \XEntropy_\alpha (P_{U|Y=y}, P_{\hat{U}|Y=y}) \Big) \Big) \Big]  \label{eq:FmeanAlphaLeak} \ \\
	& = I_{\alpha}^{\text{A}} (P_{U};P_{Y|U}). \nonumber
\end{align}
Here,
$$\XEntropy_\alpha (P_U, P_{\hat{U}}) = \frac{\alpha}{1-\alpha} \log \sum_{u} P_{U}(u) P_{\hat{U}}^{\frac{\alpha-1}{\alpha}}(u)$$
is the R\'{e}nyi cross entropy proposed in~\cite[Eq.~(7)]{Ding2024_ISIT} as a $\tilde{f}$-mean uncertainty measure incurred by $P_{\hat{U}}$ w.r.t. the probability $P_{U}$. The minimization of this R\'{e}nyi cross entropy over $P_{\hat{U}}$ gives R\'{e}nyi entropy: $\min_{P_{\hat{U}}} \XEntropy_\alpha (P_U, P_{\hat{U}}) = H_{\alpha} (P_{U})$~\cite[Theorem~1]{Ding2024_ISIT}.

Eq~\eqref{eq:FmeanAlphaLeak} in fact rewrites \cite[Definition~1]{Ding2024_ISIT} viewing $\Leak_\alpha (U \rightarrow Y)$ as a $\tilde{f}$-mean leakage measure. Based on this interpretation, we propose a $Y$-elementary leakage incurred at each channel output instance $y$ as

\begin{align}
    &\Leak_\alpha  (U \rightarrow y) \nonumber \\
    &= \min_{P_{\hat{U}}}  \XEntropy_\alpha (P_U, P_{\hat{U}} ) -  \min_{P_{\hat{U}|Y=y}} \XEntropy_\alpha (P_{U|Y=y}, P_{\hat{U}|Y=y}) \label{eq:PML1}\\
    & =
    \begin{cases}
    	-\log \frac{  \min\limits_{P_{\hat{U}|Y=y}} \max\limits_{u \in \supp(P_{U|Y=y})}  P^{-1}_{\hat{U}|Y}(u|y)  }{\min\limits_{P_{\hat{U}}} \max\limits_{u \in \supp(P_U)}  P^{-1}_{\hat{U}}(u) }&  \alpha = 0 \\
        \frac{\alpha}{\alpha-1} \log \frac{ \min\limits_{P_{\hat{U}|Y=y}} \E_{P_{U|Y=y}}\big[ P_{\hat{U}|Y}^{\frac{\alpha-1}{\alpha}}(u|y) \big]}{\min\limits_{P_{\hat{U}}} \E_{P_{U}}\big[ P_{\hat{U}}^{\frac{\alpha-1}{\alpha}}(u) \big]}  & \alpha \in (0,1) \\
        \log \frac{ \max\limits_{P_{\hat{U}|Y=y}} \prod\limits_{u} P_{\hat{U}|Y} (u|y)^{P_{U|Y}(u|y)} }{ \max\limits_{P_{\hat{U}}} \prod\limits_{u} P_{\hat{U}} (u)^{P_{U}(u)}  }
        & \alpha = 1 \\
        \frac{\alpha}{\alpha-1} \log \frac{ \max\limits_{P_{\hat{U}|Y=y}} \E_{P_{U|Y=y}}\big[ P_{\hat{U}|Y}^{\frac{\alpha-1}{\alpha}}(u|y) \big]}{\max\limits_{P_{\hat{U}}} \E_{P_{U}}\big[ P_{\hat{U}}^{\frac{\alpha-1}{\alpha}}(u) \big]}  & \alpha \in (1,\infty) \\
        \log \frac{ \max\limits_{P_{\hat{U}|Y=y}} \E_{P_{U|Y=y}}\big[ P_{\hat{U}|Y}(u|y) \big]}{\max\limits_{P_{\hat{U}}} \E_{P_{U}}\big[ P_{\hat{U}}(u) \big]}  & \alpha = \infty
    \end{cases} \label{eq:PML2} \\
    &= H_{\alpha}(P_U) - H_{\alpha}(P_{U|Y=y}) \nonumber \\
    &= \frac{1}{\alpha-1} \log \frac{\sum_{u} P_{U|Y}^{\alpha}(u|y)}{\sum_{u} P_{U}^{\alpha}(u)}, \qquad\quad \forall \alpha \in [0,\infty).
\end{align}
Eq.~\eqref{eq:PML1} quantifies the $Y$-elementary information leakage as the difference of the best prior and posterior uncertainty reduction at the adversary for each elementary channel output $y$.
In~\eqref{eq:PML2}, the cases when $\alpha = 0$ and $\alpha = 1$ can be simplified to $\log \frac{  \max\limits_{P_{\hat{U}|Y=y}} \min\limits_{u \in \supp(P_{U|Y=y})}  P_{\hat{U}|Y}(u|y)  }{\max\limits_{P_{\hat{U}}} \min\limits_{u \in \supp(P_U)}  P_{\hat{U}}(u) }$ and  $\max\limits_{P_{\hat{U}|Y=y}} \E_{P_{U|Y=y}}[\log P_{\hat{U}|Y=y} (\cdot|y)] - \max\limits_{P_{\hat{U}}} \E_{P_{U}}[\log P_{\hat{U}} (\cdot)]$, respectively. The segmented eq.~\eqref{eq:PML2} tries to express each case in terms of multiplicative/fractional posterior vs. prior gain/loss, which coincides with the idea in~\cite{Issa2020_MaxL_JOURNAL,Liao2019_AlphaLeak,Saeidian2023_PML}.
Eq.~\eqref{eq:PML2} also shows a conversion from minimization to maximization for $1$-crossing $\alpha$, similar to~\eqref{eq:ITWmissing}.
The reason is the change from uncertainty to certainty measure when $\alpha$ crosses $1$: for $\alpha>1$, $P_{\hat{U}}^{\frac{\alpha-1}{\alpha}}(u)$ measures certainty (incurred by $P_{\hat{U}}$); for $\alpha<1$, $(P_{\hat{U}}^{-1}(u))^{\frac{1-\alpha}{\alpha}}$ measures uncertainty.

\begin{proposition} \label{prop:PML}
  The maximal leakage over all attributes $U$ of $X$ for given $P_{X}$ and $P_{Y|X}$ over Markov chain $U-X-Y-\hat{U}$ is
  \begin{equation}
    \sup_{P_{U|X}}\Leak_\alpha (U \rightarrow y) = D_{\alpha}(P_{X|Y=y} \| P_{X}), \quad \forall \alpha \in [0,\infty)
  \end{equation}
\end{proposition}
\begin{IEEEproof}
    Let $P_{U_\alpha} (u) = \frac{P_U^{\alpha}(u)}{\sum_{u \in } P_U^{\alpha}(u)}$ for all $u$ and $P_{U_\alpha|Y}(u|y) = \frac{P_{Y|U}(y|u) P_{U_\alpha}(u)}{P_Y(y)}$ for all $u,y$.\footnote{Note that $P_{U_\alpha|Y}(u|y)$ is a conditional probability as for Markov chain $U-X-Y$ with fixed $P_{X}$ and $P_{Y|X}$, we have $P_{X}(x) = \sum_{u} P_{X|U}(x|u)P_{U_\alpha}(u)$ and so $\sum_{u}P_{Y|U}(y|u) P_{U_\alpha}(u) = \sum_{x}P_{Y|X}(y|x) \sum_{u} P_{X|U}(x|u) P_{U_\alpha}(u) = \sum_{x}P_{Y|X}(y|x) P_{X}(x) = P_{Y}(y)$.}
    Then,
    $\Leak_\alpha  (U \rightarrow y) = D_\alpha(P_{U_\alpha | Y=y}  \| P_{U_\alpha})$ and
    \begin{align}
        \sup_{P_{U|X}}\Leak_\alpha (U \rightarrow y)
        &= \sup_{P_{U_{\alpha}|X}} D_\alpha(P_{U_\alpha | Y=y}  \| P_{U_\alpha}) \nonumber\\
        &= \sup_{P_{U|X}} D_\alpha(P_{U | Y=y}  \| P_{U}) \label{eq:ChangeVar} \\
        &= D_\alpha(P_{X | Y=y}  \| P_{X}), \label{eq:DataProcessing}
    \end{align}
    where \eqref{eq:ChangeVar} is simply a change of the notation of the decision variable and \eqref{eq:DataProcessing} is because of the data processing inequality of R\'{e}nyi divergence~\cite[Theorem~1]{Erven2014_JOURNAL}.
\end{IEEEproof}
The so-called pointwise maximal leakage (PML)~\cite[Definition~1]{Saeidian2023_PML} first defines~\cite[Eq.~(3)]{Saeidian2023_PML}
\begin{align}
    \Leak_{\infty}(U \rightarrow y)
    &= \log \frac{ \max\limits_{P_{\hat{U}|Y=y}} \E_{P_{U|Y=y}}\big[ P_{\hat{U}|Y}(u|y) \big]}{\max\limits_{P_{\hat{U}}} \E_{P_{U}}\big[ P_{\hat{U}}(u) \big]} \\
    &=  \log \frac{ \max\limits_{P_{\hat{U}|Y=y}} \Pr(\hat{U}=U|Y=y)}{\max\limits_{P_{\hat{U}}} \Pr(\hat{U}=U)},
\end{align}
the $Y$-elementary leakage on $U$ at order $\alpha = \infty$; then takes the supremum $\sup_{P_{U|X}}\Leak_\infty (U \rightarrow y)$ and call it PML. It is then proved in~\cite[Eq.~(3)]{Saeidian2023_PML} that PML equals the $\infty$-order R\'{e}nyi divergence
$
    \sup_{P_{U|X}}\Leak_\infty (U \rightarrow y) = D_\infty (P_{X | Y=y}  \| P_{X}),
$
which is Proposition~\ref{prop:PML} in the case of $\alpha = \infty$.

\begin{remark}
    $\Leak_\alpha (U \rightarrow y)$ could be negative, which does not disqualify it from being a $Y$-elementary leakage measure, as the elementary relative information gain, e.g, $\iota_{P_{X|Y=y} \| P_{X}}$ in R\'{e}nyi divergence, is not necessarily nonnegative. However, $\sup_{P_{U|X}}\Leak_\alpha (U \rightarrow y) \geq 0$ always due to the nonnegativity of R\'{e}nyi divergence.
\end{remark}

\begin{remark}
	Proposition~\ref{prop:PML} is consistent with the existing result $I_{\alpha}^{\text{A}} (P_{U};P_{Y|U}) = I_{\alpha}^{\text{S}} (P_{U_{\alpha}};P_{Y|U}) \leq I_{\alpha}^{\text{S}} (P_{X};P_{Y|X})$~\cite{Verdu2021_ErrExp_ENTROPY}.
\end{remark}

\section{$\delta$-approximation of $\alpha$-Leakage: Tail Bound}
\label{sec:Tail}

With the $\tilde{f}$-mean expression, we are ready to propose the $\delta$-approximation of $\alpha$-leakage, similar to the idea of $(\epsilon, \delta)$-differential privacy \cite{Dwork2014_JOURNAL}. 

We first derive an $XY$-elementary leakage from $D_\alpha(P_{X|Y=y}  \|  P_X)$. The maximal relative information gain incurred by the optimal decision $P_{\hat{X}|Y}^*$ for each elementary event $(x,y)$ with the joint probability $P_{Y|X} \otimes P_X (x,y) = P_{Y|X}(y|x) P_{X}(x)$ is
\begin{align}
    \iota_{P_{\hat{X}|Y}^* \| P_{X}}(x,y)
    &= \log \frac{P_{\hat{X}|Y}^*(x|y)}{P_{X}(x)} \\
    &= \log \frac{P_{X|Y}^{\alpha}(x|y)/P_{X}^{\alpha}(x) }{\sum_{x} P_{X|Y}^{\alpha}(x|y)/P_{X}^{\alpha-1}(x) }
\end{align}
By Eq.~\eqref{eq:SibsonRenyiDiv}, the Sibson mutual information defines the moment generating function of $Y$- and $XY$-elementary leakage, respectively, as
\begin{align}
    &\exp(\frac{\alpha-1}{\alpha} I_{\alpha}^{\text{S}} (P_X, P_{Y|X})) \nonumber \\
    & \qquad\quad = \E_{P_Y} \big[ \exp \big( \frac{\alpha-1}{\alpha} D_\alpha(P_{X|Y=y}  \|  P_X) \big) \big] \label{eq:MomentY} \\
    & \qquad\quad = \E_{P_{Y|X} \otimes P_X } \big[ \exp \big( \frac{\alpha-1}{\alpha} \iota_{P_{\hat{X}|Y}^* \| P_{X}}(X,Y) \big]. \label{eq:MomentXY}
\end{align}

A tail bound on $Y$- and $XY$-elementary leakage can be directly obtained by the exponential Chebyshev's inequality: for all $\alpha \in (1,\infty)$,
\begin{align}
    & \Pr( D_\alpha(P_{X|Y=y}  \| P_X) > \epsilon ) \nonumber \\
    &\qquad\qquad\quad \le \exp(\frac{\alpha-1}{\alpha} (I_{\alpha}^{\text{S}} (P_X, P_{Y|X}) - \epsilon) ), \label{eq:TailY} \\
    & \Pr( \iota_{P_{\hat{X}|Y}^* \| P_{X}}(X,Y) > \epsilon ) \nonumber \\
    &\qquad\qquad\quad \le \exp(\frac{\alpha-1}{\alpha} (I_{\alpha}^{\text{S}} (P_X, P_{Y|X}) - \epsilon) ).  \label{eq:TailXY}
\end{align}
Following the idea of \cite[Definition~5]{Saeidian2023_PML}, the \emph{$\delta$-approximation of privacy leakage} refers to the situation when the probability of elementary leakage being greater than a positive $\epsilon$ is no greater than $(1-\delta)$.
By~\eqref{eq:TailY} and  \eqref{eq:TailXY}, enforcing
\begin{equation}
 	\exp(\frac{\alpha-1}{\alpha} (I_{\alpha}^{\text{S}} (P_X, P_{Y|X}) - \epsilon) ) \leq 1-\delta, \quad
\end{equation}
for any $\alpha \in (1,\infty)$ sufficiently achieves the $\delta$-approximation of $\epsilon$-upper bounded $\alpha$-leakage such that $ \Pr( D_\alpha(P_{X|Y=y}  \| P_X) > \epsilon )  \le 1-\delta$ and $ \Pr( \iota_{P_{\hat{X}|Y}^* \| P_{X}}(X,Y) > \epsilon )  \le 1-\delta$.

\section{Capacity of $\alpha$-leakage}
Similar to the channel capacity in communication systems, we refer to the capacity of information leakage as the maximal privacy leaked via channel $P_{Y|X}$ over all channel input $P_{X}$. With the results in Section~\ref{sec:Sibson}, we show in this section that the R\'{e}nyi capacity is an $\alpha$-leakage capacity. This result will be formally derived with a $\tilde{f}$-mean relative information interpretation of the generalized Blahurt-Arimoto algorithm in~\cite{Arimoto1976_CompErrExp} for computing the R\'{e}nyi capacity.

\subsection{Unconditional $\tilde{f}$-mean Relative Information}\label{sec:Decompose1}
For $\tilde{D}_{\alpha} \big( P_{\hat{X}|Y=y} \|  P_X  | P_{X|Y=y}  \big)$ being a conditional $\tilde{f}$-mean divergence, we define the corresponding unconditional divergence by equating the frequency probability to reference probability $P_{X|Y=y} = P_{X}$, which also removes the dependence on $Y$ in the decision variable:
\begin{align}
	\tilde{D}_{\alpha} \big( P_{\hat{X}} \|  P_X   \big) \nonumber
    &= \tilde{D}_{\alpha} \big( P_{\hat{X}} \|  P_X  | P_{X}  \big) \nonumber\\
	&= \frac{\alpha}{\alpha-1} \log \sum_{x\in\X} P_X(x)    \Big( \frac{P_{\hat{X}}(x)}{P_X(x)}  \Big)^{\frac{\alpha-1}{\alpha}}  \nonumber \\
	&  =  \frac{\alpha}{\alpha-1} \log \sum_{x\in\X} P_X^{\frac{1}{\alpha}}(x)  P_{\hat{X}}^{\frac{\alpha-1}{\alpha}}(x). \label{eq:TDVar}
\end{align}
This $\tilde{f}$-mean divergence is closely related to the R\'{e}nyi divergence: for all $\alpha \in [0,\infty)$,
\begin{equation} \label{eq:UnCondRenyi}
    \tilde{D}_{\alpha} \big( P_{\hat{X}} \|  P_X   \big) = - D_{\frac{1}{\alpha}} (P_{X} \| P_{\hat{X}}).
\end{equation}
Therefore, $\max \tilde{D}_{\alpha} \big( P_{\hat{X}} \|  P_X   \big) = - \min D_{\frac{1}{\alpha}} (P_{X} \| P_{\hat{X}})=0$.

\subsection{Decompositions by $\tilde{D}_{\alpha} \big( P_{\hat{X}} \|  P_X   \big)$}
We show below two decompositions by the proposed unconditional $\tilde{f}$-mean.
\begin{proposition} \label{prop:XentropyDecompose}
    For all $\alpha \in [0,\infty)$,
    \begin{align}
        & H_\alpha (P_X,P_{\hat{X}}) = H_{\alpha} (P_X) - \tilde{D}_{\alpha} ( P_{\hat{X}} \|  P_{X_\alpha} ) \label{eq:Decompse1} \\
        & \tilde{D}_{\alpha} \big( P_{\hat{X}|Y=y} \|  P_X  | P_{X|Y=y}  \big) = \nonumber \\
        & \qquad\quad  D_{\alpha}(P_{X|Y=y} \| P_{X}) + \tilde{D}_{\alpha} \big( P_{\hat{X}|Y=y} \| P_{\hat{X}|Y=y}^* \big)  \label{eq:Decompse2}
    \end{align}
\end{proposition}
%
Proposition~\ref{prop:XentropyDecompose} is obtained by simple algebraic works (proof omitted).
By~\eqref{eq:Decompse1}, $ \min_{P_{\hat{X}}}H_\alpha (P_X,P_{\hat{X}}) = H_{\alpha} (P_X) - \max_{P_{\hat{X}}} \tilde{D}_{\alpha} ( P_{\hat{X}} \|  P_{X_\alpha}) = H_{\alpha} (P_X)$, which is another proof of \cite[Theorem~1]{Ding2024_ISIT}.
For $\alpha = 1$, $\tilde{D}_{1} ( P_{\hat{X}} \|  P_{X} )  = - D_1(P_{X}  \|  P_{\hat{X}} )$, where $D_1(\cdot \| \cdot)$ is the Kullback-Leibler divergence and we have~\eqref{eq:Decompse1} being $H_1 (P_X,P_{\hat{X}}) = H_1 (P_X) + D_1(P_{X}  \|  P_{\hat{X}} )$, the well-known Shannon order decomposition of the cross entropy.
By~\eqref{eq:Decompse2}, $\max_{P_{\hat{X}|Y=y}} \tilde{D}_{\alpha} \big( P_{\hat{X}|Y=y} \|  P_X  | P_{X|Y=y}  \big) = D_{\alpha}(P_{X|Y=y} \| P_{X}) + \max_{P_{\hat{X}|Y=y}} \tilde{D}_{\alpha} \big( P_{\hat{X}|Y=y} \| P_{\hat{X}|Y=y}^* \big) = D_{\alpha}(P_{X|Y=y} \| P_{X})$, which provides another proof of \eqref{eq:RenyiLeakge}.

\subsection{Generalized Blahurt-Arimoto Method~\cite{Arimoto1976_CompErrExp}}
A generalized\footnote{It is shown in~\cite{Arimoto1976_CompErrExp} that for $\alpha=1$ the generalized Blahut-Arimoto method reduces to Blahut-Arimoto method \cite{Arimoto1972_TIT_BA,Blahut1972_TIT_BA}	for computing Shannon capacity. } Blahut-Arimoto method have been developed to compute R{\'e}nyi capacity~\cite{Arimoto1976_CompErrExp}.
We first demonstrate that maximizing $\tilde{f}$-mean information gain $\tilde{D}_{\alpha}  ( P_{\hat{X}|Y=y} \|  P_X  | P_{Y|X} \otimes P_X)$ over channel input $P_X$ and estimation decision $P_{\hat{X}|Y}$ results in R\'{e}nyi capacity. To do so, we may use the the decomposition in Eq.~\eqref{eq:SibsonIDnew}, derived in a similar fashion to~\eqref{eq:Decompse2} and the Sibson identity~\cite{Sibson1969_InfRadius,Nakiboglu2019_Renyi_Capcity_Centre,Verdu2021_ErrExp_ENTROPY,Polyanskiy2010_Allerton,Hayashi2017_TIT}.
We can then use the generalized Blahut-Arimoto method~\cite{Arimoto1976_CompErrExp} for maximization of $\tilde{D}_{\alpha}  ( P_{\hat{X}|Y=y} \|  P_X  | P_{Y|X} \otimes P_X)$ iteratively.

Let us restate \cite[Theorem~1-(2)]{Arimoto1976_CompErrExp} in terms of the $\tilde{f}$-mean relative information gain measure.

\begin{theorem} \label{theo:BA}
	For all $\alpha \in [0,\infty)$,
	\begin{multline} \label{eq:BANew}
		\max_{P_X} \tilde{D}_{\alpha} ( P_{\hat{X}|Y=y} \|  P_X  | P_{Y|X} \otimes P_X  \big) \\
		 = \log \sum_{x\in\X}  \Big( \sum_{y \in \Y} P_{Y|X}(y|x)  P_{\hat{X}|Y}^{\frac{\alpha-1}{\alpha}}(x|y)  \Big)^{\frac{\alpha}{\alpha-1}}
	\end{multline}
	with the maximizer
    $$P_X^*(x) = \frac{\big( \sum_{y \in \Y} P_{Y|X}(y|x) P_{\hat{X}|Y}^{\frac{\alpha-1}{\alpha}}(x|y) \big)^{\frac{\alpha}{\alpha-1}}}{\sum_{x\in\X} \big( \sum_{y\in\Y} P_{Y|X}(y|x) P_{\hat{X}|Y}^{\frac{\alpha-1}{\alpha}}(x|y) \big)^{\frac{\alpha}{\alpha-1}}},\forall x. $$
\end{theorem}
\begin{IEEEproof}
    The maximand in~\eqref{eq:BANew} is
	\begin{multline*}
        \tilde{D}_{\alpha} ( P_{\hat{X}|Y=y} \|  P_X  | P_{Y|X} \otimes P_X \big) \\
        =  \frac{\alpha}{\alpha-1}  \log \sum_{x\in\X} P_{X}^{\frac{1}{\alpha}}(x) \sum_{y\in\Y} P_{Y|X}(y|x) P_{\hat{X}|Y}^{\frac{\alpha-1}{\alpha}}(x|y).
    \end{multline*}
    With $P_X^*(x)$, we have the decomposition
    \begin{multline}\label{eq:SibsonIDnew}
        \tilde{D}_{\alpha} ( P_{\hat{X}|Y=y} \|  P_X  | P_{Y|X} \otimes P_X \big) \\
         = \tilde{D}_{\alpha} \big( P_X^* \|  P_X   \big)  + \tilde{D}_\alpha(P_{\hat{X}|Y=y} \| P_X^* | P_{Y|X} \otimes P_X^* ).
    \end{multline}
    As $\max_{P_X} \tilde{D}_{\alpha} \big( P_X^* \|  P_X   \big) = 0$, $\tilde{D}_{\alpha} ( P_{\hat{X}|Y=y} \|  P_X  | P_{Y|X} \otimes P_X  \big)$ reaches maximum at $P_X = P_X^*$
    \begin{multline*} 
    	\tilde{D}_\alpha(P_{\hat{X}|Y=y} \| P_X^* | P_{Y|X} \otimes P_X^* )  \\
    	=  \log \sum_{x}  \Big( \sum_{y} P_{Y|X}(y|x)  P_{\hat{X}|Y}^{\frac{\alpha-1}{\alpha}}(x|y)  \Big)^{\frac{\alpha}{\alpha-1}}. \quad \quad \IEEEQEDhere
    \end{multline*}
\end{IEEEproof}
Following Theorem~\ref{theo:BA}, R{\'e}nyi capacity\footnote{There are different versions of R{\'e}nyi capacity and~\eqref{eq:RenyiCap} is one of them. See~\cite[Sec.4.3]{Aishwarya2020_ArimotoCondEntropy} and \cite{Verdu2021_ErrExp_ENTROPY}.}~\cite{Csiszar1972,Arimoto1977} is
	\begin{align}
		C_{\alpha} ( P_{Y|X}) &= \max_{P_X} I_{\alpha}^{\text{S}} (P_X,P_{Y|X})   \label{eq:RenyiCap} \\
		&= \max_{P_X} \max_{ P_{\hat{X}|Y} }  \tilde{D}_{\alpha}  ( P_{\hat{X}|Y=y} \|  P_X  | P_{Y|X} \otimes P_X).  \nonumber
	\end{align}
This max-max optimization parallels the well-known max-min approach $C_{\alpha} (P_{Y|X})
	= \text{max}_{P_X} \text{min}_{Q_Y}  D_{\alpha} (P_{Y|X} \| Q_Y | P_{X})$ and immediately suggests an iterative algorithm of alternating maximizations, which is proposed in~\cite[pp.667]{Arimoto1976_CompErrExp}
and can be written as
\begin{align}
    & P_{\hat{X}|Y} \coloneqq \argmax_{P_{\hat{X}|Y}} \tilde{D}_{\alpha} \big( P_{\hat{X}|Y=y} \|  P_X  | P_{Y|X} \otimes P_X  \big),  \\
    & P_X \coloneqq \argmax_{P_X} \tilde{D}_{\alpha} \big( P_{\hat{X}|Y=y} \|  P_X  | P_{Y|X} \otimes P_X  \big).
\end{align}
The sequence $\Set{  \tilde{D}_{\alpha}  ( P_{\hat{X}|Y=y}^{(t)} \|  P_X^{(t)}  | P_{Y|X} \otimes P_X^{(t)})
\colon t = 0,1,\dotsc}$ converges monotonically to $C_{\alpha} ( P_{Y|X}) $~\cite[Th.~3]{Arimoto1976_CompErrExp}.

\section{Conclusion}
In this paper, we used $\tilde{f}$-mean $\alpha$-leakage interpretation of the Sibson mutual information to formulate $Y$- and $XY$-elementary information leakage measures. We developed a sufficient condition to achieve $\delta$-approximation of $\epsilon$-bounded $Y$- and $XY$-elementary leakage. We also proposed an alternative $Y$-elementary information leakage to generalize existing pointwise maximal leakage. We proved that R\'{e}nyi capacity captures the maximal $\alpha$-leakage over all adversary's inference decisions and channel inputs.
This enabled us to use the generalized  Blahurt-Arimoto algorithm for computing the $\tilde{f}$-mean information gain measure.

The following directions are viable for future work. The relationship between $f$-mean and $\tilde{f}$-mean measures, particularly for  $\alpha \in [0,1)$, can be further explored.
The dependence of optimal estimation decision $P_{\hat{X}|Y}^*(x|y)$ to $\alpha$ and its interpretation in terms of information leakage can also be discussed.
Parallel to \eqref{eq:RenyiLeakge}, another well known variational expression is $D_{\alpha}(P_{X|Y=y} \| P_{X}) = \frac{1}{1-\alpha} \min_{P_{\hat{X}|Y=y}} \big\{ \alpha D_1(P_{\hat{X}|Y=y} \| P_{X|Y=y}) + (1-\alpha) D_1(P_{\hat{X}|Y=y} \| P_{X})  \big\}$~\cite[Theorem~30]{Erven2014_JOURNAL}. It is interesting to see whether similar interpretations hold for information leakage as well.




\IEEEtriggeratref{11}

\bibliographystyle{IEEEtran}
\bibliography{privacyBIB_ITW2024}

\end{document}